\documentclass[aps,prl,twocolumn,nofootinbib, showpacs]{revtex4-1}

\usepackage{epsfig,amsmath}
\usepackage{graphicx}
\usepackage{dcolumn}
\usepackage{bm}
\usepackage{bbold}
\newcommand{\beq}{\begin{equation}}
\newcommand{\eeq}{\end{equation}}
\newcommand{\beqa}{\begin{eqnarray}}
\newcommand{\eeqa}{\end{eqnarray}}

\newcommand{\la}{\langle} 
\newcommand{\ra}{\rangle}

\def\nat#1{{ Nature} {\bf#1}}

\def\njp#1{{ New\ J.\ Phys.} {\bf#1}}

\def\oc#1{{ Opt.\ Commun.} {\bf#1}}
\def\ol#1{{ Opt.\ Lett.} {\bf#1}}

\def\pla#1{{ Phys.\ Lett. A\/} {\bf#1}}

\def\pra#1{{ Phys.\ Rev. A\/} {\bf#1}}

\def\prd#1{{ Phys.\ Rev. D\/} {\bf#1}}

\def\prl#1{{ Phys.\ Rev.\ Lett.} {\bf#1}}

\def\sci#1{{ Science} {\bf#1}}

\begin{document}

\title{Are Quantum Objects Born with Duality?}
\author{X.-F. Qian$^{1}$}
\email{xiaofeng.qian@rochester.edu}
\author{G. S. Agarwal$^{2}$}
\email{girish.agarwal@tamu.edu}
\affiliation{$^{1}$The Institute of Optics, and Center for Coherence and Quantum Optics, University of Rochester, Rochester, NY 14627, USA \\
$^{2}$ Institute for Quantum Science and Engineering, Department of Biological and Agricultural Engineering, and Department of Physics and Astronomy, 
Texas A\&M University, College Station, TX 77843, USA }


\date{\today }

\begin{abstract}
We study wave-particle duality by exploring for the first time effects of a quantum object's source. A single photon emitted from a pair of nonlocally entangled two-level atoms is specifically analyzed. Surprisingly, duality is found to be a conditional phenomenon depending on the photon's atomic source. It can be tuned maximum, medium, and even minimum (completely absent) by the atomic state purity through an exact quadratic relation that can be called Duality Pythagorean Theorem. The analysis shows a new way of investigating duality by accounting how the single quantum object is created. The result sheds a new light on the fundamental understanding of the completeness of wave-particle duality, and can be tested in various practical physical systems.  
\end{abstract}


\maketitle

\noindent{\bf Introduction:} Wave and particle are two coexisting fundamental aspects of phenomena of each single quantum object \cite{Broglie-23}. According to Bohr \cite{Bohr-28} and we rephrase here, the two aspects are contradictory but must be regarded as complementary in the sense that only the totality of them fully characterizes the possible information about the object. On the other hand, a quantum object is obviously also fully characterized by the totality of its various specific physical properties, e.g., coherence, position, momentum, spin, etc. Some of them, such as coherence, correspond to wave description, and some others, like location (or localized position), correspond to particle characterization. Such correspondences were first established quantitatively by Wootters and Zurek \cite{Wootters-Zurek-79} and then followed by many others \cite{Glauber-86, Greenberger-Yasin-88, Mandel-91, Jaeger-Horne-Shimony-93, Jaeger-Shimony-Vaidman-95, Englert-96} to achieve a complementarity inequality, $V^2+D^2\le1$, between single quantum object wave interference visibility $V$ and particle location distinguishability $D$. Recent studies have extended the inequality by considering uncertainty \cite{Zhu-etal-12}, degree of polarization \cite{Lahiri-11, Zela-14, EQV-17, Zela-OL18, Zela-18}, total visibility \cite{FriebergPRL}, and other coherence measures \cite{Luis-08,Luis-18}. Apparently, there are still many other specific physical properties such as momentum, spin, etc., that are not accounted by the inequality, which indicates its potential incompleteness. This has been resolved recently by taking into account entanglement (measured by concurrence $C$ \cite{Wootters}) of all remaining intrinsic properties (degrees of freedom) of the single quantum object, leading to a three-way complementary idenity, $V^2+D^2 +C^2 =1$ \cite{QVE-18, Qian-etal18, Jacob-Bergou}. 

Although quantitative connections between physical properties and wave-particle descriptions can now be regarded as complete, another layer of questions remains open. Specific physical properties, such as coherence (a wave property) and localized position (a particle property), are often determined by the source and the mechanism through which the quantum object is created. Then it is natural to ask: whether the wave-particle nature of a quantum object is also controlled by its source? If yes, how? Can duality be tuned? Are quantum objects born with duality? In this Letter, we provide the first attempt to answer these questions by analyzing in detail wave-particle duality in the context of the quantum object's source. 

\begin{figure}[h!]
\includegraphics[width=5cm]{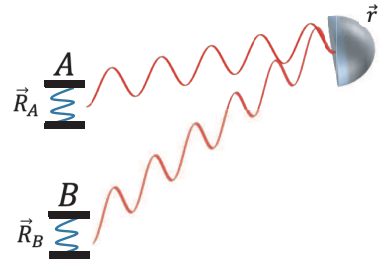}
\caption{Two-path interference. Two entangled atoms $A$ and $B$ (at $\vec{R}_A$ and  $\vec{R}_B$ respectively) emit a photon for interference detection at $\vec{r}$. } 
\label{2path}
\end{figure}

Two-path (e.g., double-slit) interference is nearly the uniform scenario employed in the analysis of duality for various quantum objects, such as electron \cite{Davisson-Germer}, atom $^{85}{\rm Rb}$ \cite{DNR98}, molecule $C_{60}$ \cite{Arndt-etal99}, surface plasmon \cite{Kolesov-etal09}, etc. Here we consider a single photon that is emitted by two nonlocally entangled identical two-level atoms, where the locations of the two distant atoms serve as two possible paths of the photon, see Fig.~\ref{2path} for illustration. Such a scenario was first experimentally realized by Eichmann et al., \cite{Eichmann-etal-93} with a laser beam exciting either one of the two trapped $^{198}{\rm Hg}^{+}$ ions to produce fluorescence single photon interference and wave-particle duality analysis, see for example detailed theoretical analyses in \cite{Huang-etal-96, Itano-etal-98}. Inversely, as proposed by Cabrillo et al. \cite{Cabrillo-etal-99}, single photon detection can on the other hand create heralded atom-atom entanglement, which was recently realized by Slodi\v cka et al. \cite{Slodicka-etal-13}, with two trapped $^{138}{\rm Ba}^{+}$ ions.

Generated by such a two-atom source, the single-photon which-way (particle) property directly corresponds to the information of which atom experiences the emission, and single photon interference (wave) property is related to the superposed atomic state. This allows us to perform a quantification of the photon's wave and particle features through parameters of the atomic source state and retrieve the conventional complementarity inequality. More importantly, detailed analysis of the source shows that the amount of information available for exchange between the photon's waveness and particleness is fully determined by the atomic state purity through an exact Pythagorean relation. Experimental configuration established by Slodi\v cka et al. \cite{Slodicka-etal-13}, provides a practically suitable platform for a test of our theoretical result, and will be discussed later. \\




\noindent{\bf Single Photon Source:} We consider a photon generated by the fluorescence of a pair of identical two-level atoms ($A,B$) in an entangled single excitation state, i.e.,
\beq 
|\psi_{AB}\ra= c_a|e_A\ra|g_B\ra+c_b|g_A\ra|e_B\ra, \label{pure-atom1}
\eeq 
where $|e\ra$, $|g\ra$ are the excited and ground states of atom $A$ or $B$. One way of creating such an entangled state is to use Coulomb repulsion between the ions with some laser couplings \cite{Cirac-Zoller-95}. Another way is to use laser pulses to excite one of two atoms from ground state to excited state, and then perform a single photon detection to generate a heralded entangled atomic state, which was indirectly \cite{Eichmann-etal-93} and directly \cite{Slodicka-etal-13} realized in experiments. Since our analysis is based on single photon interference and detection, the latter procedure is more practically suitable for an experimental realization. 

In practice, it is often impossible to prepare a pure atomic state (\ref{pure-atom1}). This may be caused by atomic thermal fluctuations, recoil due to absorption or emission of a photon by the atom, interaction with external systems and fields, etc. Therefore the two electronic energy states $|e_A\ra|g_B\ra$ and $|g_A\ra|e_B\ra$ in general should correspond to different states of the remaining degrees of freedom of the atoms as well as the states of external parties, i.e., 
\beq 
|\psi_{AB}\ra= c_a|e_A\ra|g_B\ra|m\ra+c_b|g_A\ra|e_B\ra|n\ra.\label{pure-atom2}
\eeq 
Here $|m\ra=\sum c_{m_A, m_B, m_E} |m_A\ra|m_B\ra|m_E\ra$, $|n\ra=\sum d_{n_A, n_B, n_E} |n_A\ra|n_B\ra|n_E\ra$ represent two sets of states for all remaining degrees of freedom of atoms $A$ and $B$, as well as external parties $E$ indicated respectively by the subscripts.

By tracing out the states $|m\ra$ and $|n\ra$, i.e., all other degrees of freedom and systems, the atomic electronic energy state can be written as 
\beqa
\label{mixed}
\rho_{AB}= \left(\begin{matrix}
p_{a}  & \gamma  \\
\gamma*   & p_{b}   \\
\end{matrix} \right)
\eeqa
in the basis set $|e_A\ra|g_B\ra$, $|g_A\ra|e_B\ra$. It is noted that state (\ref{mixed}) derived from (\ref{pure-atom2}) is in fact the most general form of a mixed state. Here $p_a=|c_a|^2$ is the probability of atom $A$ in excited state and atom $B$ in ground state, i.e., $|e_A\ra|g_B\ra$ (similar for $p_b$), and $\gamma=|\gamma|e^{i\varphi}=\la m|n\ra$ is the overlap of the two collective states $|m\ra$, $|n\ra$ which is restricted by the Cauchy-Schwarz inequality $|\gamma|\le|c_ac_b|$. 


Due to spontaneous emission, a photon will be generated by either one of the atoms $A, B$. To characterize the photon properties in terms of the source parameters, one can adopt the approach taken in Refs.~\cite{Agarwal74,Agarwal11}. The total single photon field (positive frequency part) is the sum of two fields and is given as
\beqa
E^{(+)}_{AB}&=&e^{-i(k\hat{r}\cdot \vec{R}_A+\phi_A)}s_A+e^{-i(k\hat{r}\cdot \vec{R}_B+\phi_B)}s_B,
\eeqa
where $k = 2\pi/\lambda=\omega/c$ denotes the wave number of the photon generated by the two atoms and $\hat{r}=\vec{r}/|\vec{r}|$ is the unit vector in the direction of the detector. The initial phases of the atomic sources $A,B$ are denoted by $\phi_A$ and  $\phi_B$ respectively, and $s_A=|g_A\ra\la e_A|$ is the lowering operator of the two-level atom $A$ (similar for $s_B$). 

The detection of a photon, collecting all the data when registers only one photon at a time at the detector $D$, is described with probability $p_D$ that is characterized by various parameters of the atomic source, i.e., 
\beqa
p_D&=&Tr[(s_A^{+}+e^{-i\theta}s_B^{+})(s_A^{-}+e^{i\theta}s_B^{-})\rho_{AB}] \notag\\
&=&p_a+p_b+2|\gamma|\cos(\theta+\varphi),    \label{probability}
\eeqa
where we have extracted and omitted the non-relevant global phase and kept the relative one $\theta=k\hat{r}\cdot (\vec{R}_B-\vec{R}_A)+\phi_B-\phi_A$. \\



\noindent{\bf Duality Under Control:} Now it is ready to quantify wave and particle properties of the generated photon. The standard measure of wave feature is interference visibility related to the superposed atomic state (\ref{pure-atom2}) or (\ref{mixed}). It can be obtained directly from (\ref{probability}), and is given as 
\beq \label{V}
V=\frac{p_D^{max}-p_D^{min}}{p_D^{max}+p_D^{min}}=2|\gamma|.
\eeq

The particle nature of the photon is embodied by the degree to which it is localized, which in this case means to what degree it is emitted from only one of the two atoms. The standard measure is distinguishability, which is represented by the probability difference of the photon being emitted from atoms $A$ and $B$, i.e., 
\beq \label{D}
D=\frac{|p_a-p_b|}{p_a+p_b}=|p_a-p_b|.
\eeq

Due to the fact that $|\gamma| \le \sqrt{p_ap_b}$, it is straightforward to reach the conventional duality inequality
\beq \label{VD}
V^2 + D^2= (p_a-p_b)^2 + 4|\gamma|^2 \le1.
\eeq

As pointed out in \cite{QVE-18}, such an inequality is incomplete to represent Bohr's complementarity principle for it embodies neither exclusiveness nor completeness (through $V$ and $D$), two characteristic features of complementarity. Here, it is important to note that the controlling parameters $p_a,p_b, \gamma$ of the duality sum $V^2+D^2$ correspond only to properties of the atomic source. This indicates that a resolution of the incompleteness of the above inequality (\ref{VD}) needs also to trace back to the source. 

A further analysis of the sum shows that it corresponds solely to the atomic density matrix (\ref{mixed}) in the following compact way
\beq \label{VD}
V^2 + D^2= 1-4{\rm det}\rho_{AB}=2{\rm Tr}\rho_{AB}^2-1.
\eeq
One notes that ${\rm Tr}\rho_{AB}^2$ is the usual state purity measure for (\ref{mixed}), varying from $1/2$ to 1, and $2{\rm Tr}\rho_{AB}^2-1$ is one form of its normalization. For symmetry considerations and without loss of generality, one can conveniently define a normalized purity as
\beq \label{mu}
\mu_S=\sqrt{2{\rm Tr}\rho_{AB}^2-1}.
\eeq

This immediately allows to arrive at the central result of the Letter, i.e., the duality Pythagorean theorem
\beqa \label{VDmu}
V^2+D^2= \mu_S^2.
\eeqa
It shows that wave-particle duality of a photon is under control by its atomic source, i.e., the complementary behavior of $V$ and $D$ is determined by the source purity $\mu_S$. When the source state is maximally mixed ($\mu_S=0$), the generated photon can display no duality properties at all, i.e., $V=D=0$. When the source is pure ($\mu=1$), the photon can have full wave-particle duality with complete waveness ($V=1$) and complete particleness ($D=1$) both reachable.

It is important to note that interesting investigations of the value of the duality sum $V^2+D^2$ have been carried out previously\footnote{Jacob and Bergou \cite{Jacob-Bergou} explored connections of distinguishability to visibilities in two-particle interferometers, and noticed the basis invariant property of the sum.}. Recently, a similar Pythagorean relation called Polarization Coherence Theorem (PCT) was obtained by Eberly et al.~\cite{EQV-17} demonstrating connections of the duality sum to the generic degree of polarization within a single classical optical field. The PCT was then extended to generalized two-state distance measures by De Zela \cite{Zela-18}. Another important quadratic connection among visibility, coherence, and phase statistics was reported by Luis and coworkers \cite{Luis-08}. Here in our case, the normalized purity $\mu_S$ relates only to the source of the single photon. So the above result (\ref{VDmu}) shows for the first time connection between a quantum object's wave-particle duality to its source state. 

\begin{figure}[t!]
\includegraphics[width=6cm]{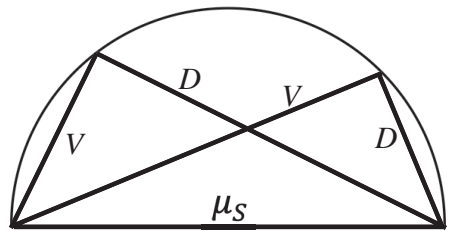}
\caption{Geometric illustration of wave-particle duality Pythagorean theorem.} 
\label{r-triangle}
\end{figure}

The quadratic form of measures in (\ref{VDmu}) indicates its direct connection to the Pythagorean relation. Therefore it can be represented geometrically with right triangles, where the value of $\mu_S$ is represented by the length of its longest side and the values of $V$, $D$ are represented by the two shorter sides. Fig.~\ref{r-triangle} illustrates schematically two examples of such right triangles, who share their longest side (which forms the diameter of a circle) and two corresponding apexes. Both $V$ and $D$ can vary between 0 and $\mu_S$, which exhausts all points on the half circle as the right angle apex. \\

\noindent{\bf Generalized Result:} We now extend the analysis to general and practical situations when additional degrees of freedom of the photon is needed. We first include considerations of vector fields where polarization degree of freedom matters. When two atoms are not perfectly identical, e.g., transition between different nearby energy levels, they may emit photons with different polarizations but still with approximately same energy and spontaneous emission decay rate. Then the description of a photon arrived at the detector needs to be modified as  
\beqa \label{pDgeneral}
p_D&=& Tr[(\hat{\epsilon}^*_a s_a^{+}+e^{-i\theta}\hat{\epsilon}^*_b s_b^{+})(\hat{\epsilon}_a s_a^{-}+e^{i\theta} \hat{\epsilon}_b s_b^{-})\rho_{AB}] \notag \\
&=& p_a+p_b+ 2|\gamma \eta| \cos(\theta+\varphi+\Delta), \label{pp}
\eeqa
where $\hat{\epsilon}_a,  \hat{\epsilon}_b$ are two respective polarization states of the spontaneously emitted photon from atoms $A,B$, and they have a generic overlap relation $\hat{\epsilon}^*_a \cdot  \hat{\epsilon}_b=\eta=|\eta|e^{i\Delta}$. Such a characterization is also consistent with traditional optical polarization analysis \cite{Wolfbook, Lahiri-etal}, where a pair of polarizers are used to introduce different polarizations of light coming from two atoms respectively. 

When polarization states are involved, polarization modulated total visibility $V_P$, proposed by Friberg and coworkers \cite{Frieberg-early,FriebergPRL}, is usually employed to represent full coherence of an optical field. It is defined as 
\beq \label{Vpola}
V_P=\sqrt{\frac{1}{2}\sum_{j=0}^{3}V_j^2}, \quad \text{with} \quad  V_j=\frac{S_j^{max}-S_j^{min}}{S_0^{max}+S_0^{min}}.
\eeq
Here $V_j$, $j=0,1,2,3$, are called polarization visibilities \cite{Frieberg-early}, and $S_j$ are four corresponding conventional polarization Stokes parameters that can be obtained as $S_j=\la \hat{S}_j \ra$ with $\hat{S}_j$ being the Stokes operators that are analogous to Pauli matrices \cite{James-etal}. Recently this modulated total polarization visibility was employed as a measure of wave property in the complementarity analysis of a quantized vector field (vector single photon) \cite{FriebergPRL}. 

Given the contribution of polarization states, we also adopt this total visibility $V_P$ to measure waveness of the photon. For the physical context here, each individual polarization visibility can be obtained as 
\beq \label{Vj}
V_0=V_1=2|\gamma\eta|, \quad V_2=V_3=2|\gamma|\sqrt{1-|\eta|^2},
\eeq
which leads to the total polarization modulated visibility 
\beq \label{VP}
V_P =2|\gamma|.
\eeq

The photon distinguishability $D$ and atomic state purity $\mu_S$ remain the same as in (\ref{D}) and (\ref{mu}). Then for the more general and practical case of unbalanced polarizations, one obtains the same form of the above duality Pythagorean relation, i.e.,  
\beq \label{VDpola}
V_P^2 + D^2=\mu_S^2.
\eeq
Since $\mu_S^2\le1$, this relation explains from a new perspective the key relation $V_P^2 + D^2\le1$ obtained in \cite{FriebergPRL}.

In more general and practical cases, non-perfectness of the atomic source or influence of external fields and parties may lead to unbalance of photon states in degrees of freedom other than polarization. Then one can always group the affected one or more degrees of freedom together and represent with a single (discrete or continuous) state $|\phi_a\ra$ indicating photon emitted by atom $A$, and $|\phi_b\ra$ for atom $B$. Then the probability of photon detection can be described in general as in (\ref{pDgeneral}) by replacing $\hat{\epsilon}_a$, $\hat{\epsilon}_b$ with $|\phi_a\ra$, $|\phi_b\ra$ respectively. It is important to note that $|\phi_a\ra$, $|\phi_b\ra$ live in an effective two dimensional space just as the polarization states $\hat{\epsilon}_a$, $\hat{\epsilon}_b$ do. It is spanned by the basis $\{|\phi_a\ra, |\bar{\phi}_a\ra\}$, where $\la\phi_a|\bar{\phi}_a\ra=0$ and $|\phi_b\ra$ can always expressed as $|\phi_b\ra=\eta|\phi_a\ra+\sqrt{1-|\eta|^2}|\bar{\phi}_a\ra$. Therefore, one can define a general total modulated visibility $V_T$ as in (\ref{Vpola}) by replacing the polarization Stokes parameters with generic two-dimensional Stokes-like parameters (see a systematic analysis of such parameters by James, et al., in Ref.~\cite{James-etal}). This further extends the polarization modulated result (\ref{VDpola}) to more generalized situations. \\

\noindent{\bf Summary:} In summary, we have investigated for the first time a photon's wave-particle duality in connection with its atomic source. The amount of information available to be exchanged between waveness and particleness is tunable and determined by the source that gives birth to the quantum object. A general duality Pythagorean theorem is obtained showing exact quantitative restrictions. Our result opens a new way of investigating and understanding duality through the perspective of a quantum object's source . 

The consideration of generic mixed state, through purity $\mu_S$, of the atomic source is a practical treatment compatible with experimental conditions.  As pointed out by Slodi\v cka et al. \cite{Slodicka-etal-13}, for their experimental setup, approximately 38\% of the incoherence (or mixedness) of the state (\ref{pure-atom1}) comes from imperfect populations, collective magnetic field fluctuations, atomic motion, atomic recoil, etc. Therefore by tuning some of the properties through cooling or other measures \cite{Slodicka-etal-12}, one is able to directly observe the duality Pythagorean theorem (\ref{VDmu}). 

It is worth emphasizing that the overall analysis of this Letter is not limited to photons. It applies to a generic single quantum object that is generated by a two-center source. For example, ``Young-type" electron interference was observed in charged-particle-impact ionization of diatomic ${\rm H}_2$ molecules, due to superposition of ionization amplitudes associated with the two hydrogen atoms \cite{Madison06}. Also, ``spontaneous emitting" an atom was achieved with optically trapped Bose-Einstein condensate (BEC) \cite{Schneble18}, and two-path interference was observed with atoms from two BECs \cite{Ketterle97}. Our analysis also has important implications in multi-path interference of a single quantum object \cite{Bagan-etal} in connection with multi-center source properties such as superradiant and subradiant emissions of multiple entangled atoms \cite{Wiegner-etal}.  \\

\noindent{\bf Acknowledgement:} XFQ acknowledges conversations with Joseph Eberly, Nick Vamivakas, Peter Milonni  and financial support from ARO W911NF-16-1-0162, ONR N00014-14-1-0260, and NSF grants PHY-1203931, PHY-1505189, and INSPIRE PHY-1539859. GSA acknowledges the support from AFOSR FA9550-18-1-0141.\\

\end{document}